\documentclass[%
 reprint,
nofootinbib,
 amsmath,amssymb,
 aps,
]{revtex4-1}

\usepackage{graphicx}
\usepackage[latin9]{inputenc}
\usepackage{amsmath}
\usepackage{amssymb}
\makeatletter
\renewcommand\[{\begin{equation}}
\renewcommand\]{\end{equation}}

\newcommand{\be}{\begin{equation}}
\newcommand{\ee}{\end{equation}}

\begin{document}

\title{Non-locality and time-dependent boundary conditions:\\
	a Klein-Gordon perspective}
\author{S. Colin}
\author{A. Matzkin}
\affiliation{                    
	Laboratoire de Physique Th\'eorique et Mod\'elisation, CNRS Unit\'e 8089,  
	CY Cergy Paris Universit\'e, \mbox{95302 Cergy-Pontoise cedex, France}\\
}

%\pacs{03.65.-w}{Quantum mechanics}
%\pacs{03.65.Pm}{Relativistic wave equations}
%\pacs{03.65.Ud}{Quantum nonlocality}

\begin{abstract}
The dynamics of a particle in an expanding cavity is investigated in the Klein-Gordon framework in a regime in which the single particle picture remains valid. The cavity expansion  represents a time-dependent boundary condition for the relativistic wavefunction. We show that this expansion induces a non-local effect on the current density throughout the cavity. Our results indicate that a relativistic treatment still contains apparently spurious effects traditionally associated with the unbounded velocities inherent to non-relativistic solutions obtained from the Schr\"{o}dinger equation. Possible reasons for this behaviour are discussed. 
\end{abstract}

\makeatother

\maketitle

\section{Introduction}

Every prediction we can make in a quantum system is encoded in the
quantum state, which is generally not localized, but extended over
all the available space. The quantum state evolves deterministically
by way of an evolution equation: the Schr\"odinger equation in non-relativistic
quantum mechanics, the Klein-Gordon or Dirac equations in single particle
relativistic quantum mechanics. Because the quantum state is extended
all over space, it has features that link together instantaneously
properties in different regions of space. However this non-locality
cannot be used to communicate. If signaling were possible, we could
communicate with the future and modify the past. It turns out however
that the Schr\"odinger equation displays in some cases signaling \cite{calderon,ximenes}.
This is not necessarily worrying because the Schr\"odinger equation
being non-relativistic there is no bound on the velocity associated with its solutions.
But sometimes it is not obvious to assert whether this form of signaling
is an artifact (employing a non-relativistic equation), or is a genuine
feature of quantum non-locality.

A concrete example was investigated in \cite{mmw18} where the authors
considered the case of a non-relativistic particle placed in a one-dimensional 
cavity whose right wall can be set in uniform motion. This system
-- an infinite well with a moving wall -- has often been suspected
of displaying some form of single-particle non-locality \cite{greenberger,mousavi,jpasp,phase}
and has very recently been found to be relevant in atomic \cite{wang,duffin}
or neutron \cite{abele19} spectroscopy. Assume a particle is prepared in
a given state of a cavity for which the right wall
can either be static or in motion (the cavity is then in expansion). At $t=0$, 
the length of the cavity is $L(t=0)=L_{0}$. The current
$j(t,x)$ at $x=\epsilon$, where $\epsilon$ lies near
the left wall (the side opposite to the wall that is allowed to move) will be different in both cases. As shown in \cite{mmw18},
the current can in principle be measured by performing
a weak measurement (as will be recalled below). It was found that the current
value is non-local, in the sense that the current difference $\Delta j$
between both cases obeys $\Delta j(t,\epsilon)\neq0$
for $t<\frac{|L_{0}-\epsilon|}{c}$ where $c$ is the 
light velocity. This can lead to a protocol based on monitoring the
current by which it is possible to send information faster-than-light.
However the results reported in \cite{mmw18} were obtained with
solutions of the Schr\"odinger equation. Indeed, solutions of the Schr\"odinger
equation with a moving wall have been known analytically for some
time for specific wall motions \cite{doescher,makowski}. But since
the Schr\"odinger equation is non-relativistic, any expansion over the
energy eigenstates will formally include states associated with arbitrarily high
energies, corresponding to supraluminal velocities. The question is
therefore whether the supraluminal results constitute an artifact
of the description of the system by the Schr\"odinger equation.

The goal of the present article is precisely to answer this question
by investigating the same system in a relativistic setting based on
the Klein-Gordon (KG) equation. We will rely on the solutions of the KG 
equation for a particle inside a uniformly expanding cavity that were
recently obtained \cite{koehn,bb,hade}. Note that several properties of 
single-particle relativistic wave equations have been investigated recently \cite{rwe1,rwe2,rwe3,rwe4}. As is well-known, the KG equation is not free of interpretation problems, but we will
consider a regime in which these problems do not appear.

The structure of the paper is as follows. We first recall the solutions
for the KG equation with a boundary condition expanding linearly in
time that were obtained recently \cite{koehn,bb,hade}. We also derive expressions for the current density. We show that the relativistic
wavefunctions and current density reduce to the non-relativistic solutions used in \cite{mmw18}
in the limit where the wall is slowly expanding and the length of
the cavity is large with respect to the Compton wavelength of the
particle. We then compare the currents as a function of time when the outer wall is moving or remains fixed. We will see that this current change
appears instantaneously inside the cavity. 
We examine whether artifacts could account for these results and discuss how a transient regime could be included in the model. 
We finally present our conclusions concerning single-particle
non-locality in the relativistic case.

\section{KG equation with a moving boundary}

\subsection{Analytical solutions for a KG particle inside a linearly
expanding cavity}

A KG particle of mass $m$ is initially trapped in a fixed
infinite well (the potential is equal to zero for $x\in[0,L_{0}]$
and to $+\infty$ elsewhere). The solutions of the KG equation 
\begin{equation}
\frac{\partial_{t}^{2}\Phi}{c^{2}}-\partial_{x}^{2}\Phi+\frac{m^{2}c^{2}}{\hbar^{2}}\Phi=0
\end{equation}
are given by
\begin{equation}
\Phi_{\pm,n}(t,x)=\frac{1}{\sqrt{2E_{n}L_{0}}}{\displaystyle e^{i\pm\frac{E_{n}t}{\hbar}}\sin(\frac{p_{n}x}{\hbar})}\label{kgfw}
\end{equation}
with $n=1,2,3,\ldots$, $p_{n}=\frac{\hbar n\pi}{L_{0}}$ and $E_{n}=\sqrt{m^{2}c^{4}+p_{n}^{2}c^{2}}$. The $+/-$ index refers to the signs in
the exponent, opposite to the sign of the energies  ($+$/$-$ refers to anti-particles/particles respectively).
These solutions are orthonormal with respect to the KG scalar product \footnote{Recall (see eg C\.{h}\ 1 of Ref.\ \cite{greiner}) that since the KG equation
	is relativistic the continuity equation $\partial_{\mu}j^{\mu}=0$ involves the
	4-vector $
	j^{\mu}\equiv\frac{i\hbar}{2m}\left(  \psi^{\ast}\partial^{\mu}\psi
	-\psi\partial^{\mu}\psi^{\ast}\right) $
	whose time component is the density $j^{0}=\rho c$ while the space component
	is the current familiar from non-relativistic quantum mechanics; we set $\left( x^{\mu }\right) =\left( ct,\mathbf{x}\right) $ for $\mu =0$ and $\mu =1,2,3$ resp. and $x^{\mu }x_{\mu }=c^{2}t^{2}-\mathbf{x}^{2}$. Hence
	normalization of the density $\int dx\rho(x)$ implies the integrand is
	$\psi^{\ast}\partial_{t}\psi-\psi\partial_{t}\psi^{\ast}.\ $This is an
	indication that the scalar product is given by Eq. (\ref{kgsp}).\ The presence
	of the time derivative is due to the fact that the KG\ equation is second
	order in time (for a full account see
	\cite{mostaKG}).}
\begin{equation}
(\Phi,\Xi)_{KG}=\int dx(\Phi^{\ast}i\hbar\partial_{t}\Xi-i\hbar(\partial_{t}\Phi^{\ast})\Xi)\, \label{kgsp}
\end{equation}
the KG probability density for a given state $\Phi$ being defined
as 
\begin{equation}
\rho_{KG}(t,x)=(\Phi^{\ast}i\hbar\partial_{t}\Phi-i\hbar(\partial_{t}\Phi^{\ast})\Phi)\,.\label{kgdens}
\end{equation}
At $t=0$, the right wall starts to move with constant velocity $v=\beta c$
and its position at time $t$ is therefore given by $L=L_{0}+vt$
(the left wall remains fixed). The analytical solutions of this problem
(a KG particle in an infinite square-well potential with a linearly expanding wall) were obtained by Koehn (see eq. 11 in \cite{koehn}); other
authors proposed a generalization shortly after \cite{bb}, and gave
an alternative method in \cite{hade}. These solutions can be written as linear
superpositions of 
\begin{equation}
\Psi_{J,n}=N_{J,n}J_{ik_{n}}(\frac{\sqrt{L^{2}-\beta^{2}x^{2}}}{\lambda_{C}\beta})\sin\left(\frac{k_{n}}{2}\ln\left(\frac{L+\beta x}{L-\beta x}\right)\right)\label{jsol}
\end{equation}
and 
\begin{equation}
\Psi_{Y,n}=N_{Y,n}Y_{ik_{n}}(\frac{\sqrt{L^{2}-\beta^{2}x^{2}}}{\lambda_{C}\beta})\sin\left(\frac{k_{n}}{2}\ln\left(\frac{L+\beta x}{L-\beta x}\right)\right)\label{ysol}
\end{equation}
where $\lambda_{C}=\frac{\hbar}{mc}$ (the Compton wavelength of the
KG particle), $k_{n}=\frac{2n\pi}{\ln\left(\frac{1+\beta}{1-\beta}\right)}$
with $n=1,2,3,\ldots$, $N_{J,n}$ and $N_{Y,n}$ being normalization
constants. $J$ and $Y$ are Bessel functions respectively regular and irregular at the origin \cite{olver1974}. Alternatively, one can use the basis of solutions 
\begin{equation}
\Psi_{-,n}=N_{J,n}J_{-ik_{n}}(\frac{\sqrt{L^{2}-\beta^{2}x^{2}}}{\lambda_{C}\beta})\sin\left(\frac{k_{n}}{2}\ln\left(\frac{L+\beta x}{L-\beta x}\right)\right)\label{psim}
\end{equation}
and 
\begin{equation}
\Psi_{+,n}=N_{J,n}J_{ik_{n}}(\frac{\sqrt{L^{2}-\beta^{2}x^{2}}}{\lambda_{C}\beta})\sin\left(\frac{k_{n}}{2}\ln\left(\frac{L+\beta x}{L-\beta x}\right)\right)\,,\label{psip}
\end{equation}
with $\Psi_{+,n}=\Psi_{J,n}$.

\subsection{The non-relativistic limit}

We will now assume $\beta\ll1$ and $\beta x\ll L\text{. }$Indeed
typically (in particular if we have experiments in mind) the wall
motion will be non-relativistic, and recall, as mentioned in the Introduction,
that we will be interested in values of the current density near the
fixed wall. Thus the orders of the Bessel functions of interest will
be very large in magnitude, with $k_{n}\approx\frac{n\pi}{\beta}$,
and the sine part of the above solutions can be approximated as 
\begin{equation}
\sin\left(\frac{k_{n}}{2}\ln\left(\frac{L+\beta x}{L-\beta x}\right)\right)\approx\sin\left(k_{n}\beta\frac{x}{L}\right)\approx\sin(n\pi\frac{x}{L})\,.
\end{equation}
To keep the expressions short in the following, we will use 
\begin{equation}
\phi_{n}=\frac{k_{n}}{2}\ln\left(\frac{L+\beta x}{L-\beta x}\right)\,.\label{phin}
\end{equation}
The argument of the Bessel functions, denoted $z$ for short, 
\begin{equation}
z\equiv\frac{\sqrt{L^{2}-\beta^{2}x^{2}}}{\lambda_{C}\beta}\,,\label{z}
\end{equation}
can also be very large (firstly $\beta$ is small, secondly $L_{0}\gg\lambda_{C}\text{)}$.
Therefore we will consider the approximation for the Bessel functions
of imaginary order in the limit where the argument is large and positive. 

For that purpose, the functions $\tilde{J}_{\nu}(z)$ and $\tilde{Y}_{\nu}(z)$
(with real $\nu$ and positive $z$) are introduced (see \cite{chapman}
section 3 for a starting point, then \cite{dunster} and \cite{olver1974})
as
\begin{equation}
\tilde{J}_{\nu}=\mathrm{sech}(\frac{\pi}{2}\nu)\mathfrak{R}(J_{i\nu}),\quad\tilde{Y}_{\nu}=\mathrm{sech}(\frac{\pi}{2}\nu)\mathfrak{R}(Y_{i\nu})\,.\label{asex}
\end{equation}
Thanks to the relations 
\begin{equation}
\mathfrak{R}(Y_{i\nu})=\mathrm{coth}(\frac{\pi}{2}\nu)\mathfrak{I}(J_{i\nu})\,,\quad\mathfrak{I}(Y_{i\nu})=-\mathrm{tanh}(\frac{\pi}{2}\nu)\mathfrak{R}(J_{i\nu})\,,
\end{equation}
the functions $J$ and $Y$ can be expressed as 
\begin{equation}
J_{i\nu}(z)=\mathrm{cosh}(\frac{\pi}{2}\nu)\tilde{J}_{\nu}(z)+i\,\mathrm{sinh}(\frac{\pi}{2}\nu)\tilde{Y}_{\nu}(z)\label{jrel}
\end{equation}
and 
\begin{equation}
Y_{i\nu}(z)=\mathrm{cosh}(\frac{\pi}{2}\nu)\tilde{Y}_{\nu}(z)-i\,\mathrm{sinh}(\frac{\pi}{2}\nu)\tilde{J}_{\nu}(z)\,.\label{yrel}
\end{equation}
Employing the asymptotic expansions given in \cite{dunster} (see
eqs. (3.17) and (3.18) of \cite{dunster}; details are given in the
Appendix) to second order, we obtain 
\begin{align}
\tilde{J}_{\nu}(z)\approx & \sqrt{\frac{2}{\pi z}}\left(\cos(z-\frac{\pi}{4})+\frac{4\nu^{2}+1}{8z}\sin(z-\frac{\pi}{4})\right)\,,\label{approx1a}\\
\tilde{Y}_{\nu}(z)\approx & \sqrt{\frac{2}{\pi z}}\left(\sin(z-\frac{\pi}{4})-\frac{4\nu^{2}+1}{8z}\cos(z-\frac{\pi}{4})\right)\,.\label{approx1b}
\end{align}
Using these forms in Eq. (\ref{jrel}), we obtain the following expressions
\begin{align}
J_{i\nu}(z) & \approx\mathrm{cosh}(\frac{\pi}{2}\nu)\sqrt{\frac{2}{\pi z}}\left(e^{i(z-\frac{\pi}{4})}[1-i\frac{4\nu^{2}+1}{8z}]\right)\nonumber \\
 & \approx\mathrm{cosh}(\frac{\pi}{2}\nu)\sqrt{\frac{2}{\pi z}}\left(e^{i(z-\frac{4\nu^{2}+1}{8z}-\frac{\pi}{4})}\right)\,,\label{approx2} & 
\end{align}
and the analytical solutions (Eqs. (\ref{psip}) and (\ref{psim})) become
\begin{equation}
\Psi_{+,n}\approx N_{+,n}\mathrm{cosh}(\frac{\pi}{2}{k_n})\sqrt{\frac{2}{\pi z}}e^{i(z-\frac{4{k_n}^{2}+1}{8z}-\frac{\pi}{4})}\sin(\phi_{n})\label{psinr1}
\end{equation}
and 
\begin{equation}
\Psi_{-,n}\approx N_{-,n}\mathrm{cosh}(\frac{\pi}{2}{k_n})\sqrt{\frac{2}{\pi z}}e^{-i(z-\frac{4{k_n}^{2}+1}{8z}-\frac{\pi}{4})}\sin(\phi_{n})\,.\label{psinr}
\end{equation}

Now let us show that the last expression (\ref{psinr}), corresponding to particles (positive energies) indeed reduces
to the solution of the Schr\"odinger equation given by \cite{makowski,mmw18}
\begin{equation}
\psi_{n}(t,x)=\sqrt{\frac{2}{L}}\exp(-\frac{i\pi^{2}\hbar n^{2}t}{2mL_{0}L})\exp(\frac{imvx^{2}}{2\hbar L})\sin(\frac{n\pi x}{L})\,.\label{psinr0}
\end{equation}
First we note that the factor $\sqrt{\frac{2}{\pi z}}\approx\sqrt{\frac{2}{\pi L}}$
when $\beta\ll1$ (justifying the factor $\sqrt{\frac{2}{L}}$ in
(\ref{psinr0}).) Secondly we have already pointed out that $\sin(\phi_{n})\approx\sin(\frac{n\pi x}{L})$
(hence the presence of $\sin(\frac{n\pi x}{L})$ in (\ref{psinr0})).
Finally the imaginary exponential term $z-\frac{4\nu^{2}+1}{8z}$
becomes in that limit $\frac{mc^{2}t}{\hbar}-\frac{mvx^{2}}{2\hbar L}+\frac{\hbar n^{2}\pi^{2}t}{2mL_{0}L}$
(the proof is given in the Appendix). 

\subsection{Normalization}

We want to compute the KG density (\ref{kgdens}) for the state 
\begin{equation}
\Psi_{-,n}(t,x)=N_{-,n}J_{-ik_{n}}(z)\sin(\phi_{n})\,.
\end{equation}
We first note that the following relation 
\begin{equation}
\frac{dJ_{-i\nu}(z)}{dz}=\frac{1}{2}(J_{-i\nu-1}(z)-J_{-i\nu+1}(z))\label{recu}
\end{equation}
holds although the orders of the Bessel functions are imaginary. Then,
from the definition of the density (\ref{kgdens}), the previous relation
and the following ones 
\begin{eqnarray}
\partial_{t}z=\frac{c}{\lambda_{C}}\frac{L}{\sqrt{L^{2}-\beta^{2}x^{2}}}=\frac{cL}{\beta\lambda_{C}^{2}z}\,,\nonumber \\
\partial_{t}\phi_{n}=\frac{-k_{n}\beta vx}{L^{2}-\beta^{2}x^{2}}=-\frac{k_{n}cx}{\lambda_{C}^{2}z^{2}}\,,\label{derivst}
\end{eqnarray}
obtained from (\ref{z}) and (\ref{phin}), we find the density 
\begin{equation}
\small{
\rho_{KG}(t,x)=N_{-,n}^{2}\hbar\sin^{2}(\phi_{n})\frac{\partial z}{\partial t}\left[2\mathfrak{Im}(J_{-ik_{n}}(z)\frac{dJ_{ik_{n}}(z)}{dz})\right]\,.}
\end{equation}
In the non-relativistic limit, the above expression can be further
approximated to 
\begin{equation}
\small{
\rho_{KG}(t,x)=N_{-,n}^{2}\hbar\sin^{2}(\phi_{n})\left(\frac{cL}{\beta\lambda_{C}^{2}z}\right)\left[\frac{4}{\pi z}\left(\frac{1}{\cosh\frac{\pi k_{n}}{2}}\right)^{2}\right]\,.}
\end{equation}
Using $z\approx\beta^{-1}\lambda_{C}^{-1}L$, we find that
\begin{equation}
\int dx\rho_{KG}(t,x)=N_{-,n}^{2}\frac{2\hbar c\beta}{\pi}\left(\frac{1}{\cosh\frac{\pi k_{n}}{2}}\right)^{2}\,.
\end{equation}
 Therefore, if we introduce the $C_{\pm,n}$ thanks to $N_{\pm,n}=C_{\pm,n}\sqrt{\frac{\pi}{2\beta c\hbar}}\frac{1}{\cosh(\frac{\pi k_{n}}{2})}$,
the normalized states will read 
\begin{align}
\Psi_{\pm,n}(t,x)=C_{\pm,n}\sqrt{\frac{\pi}{2\beta c\hbar}}\frac{1}{\cosh(\frac{\pi k_{n}}{2})}J_{\pm ik_{n}}(z)\sin\left(\phi_{n}\right)\,\label{psinewdef}
\end{align}
and we will have that $C_{\pm,n}\approx1$ in the non-relativistic
limit. 

\subsection{Current density}

The KG current density for a given state $\Phi$ is defined as 
\begin{equation}
j_{KG}(t,x)=-\hbar c^{2}(\Phi^{\ast}i\partial_{x}\Phi-i(\partial_{x}\Phi^{\ast})\Phi)\,.\label{kgcurr}
\end{equation}
Its computation for the state $\Psi_{-,n}(t,x)$ is not much different
from that of the KG probability density. First we have that 
\begin{align}
\partial_{x}z=-\frac{1}{\lambda_{C}}\frac{\beta x}{\sqrt{L^{2}-\beta^{2}x^{2}}}=-\frac{x}{\lambda_{C}^{2}z}\,,\nonumber \\
\partial_{x}\phi_{n}=\frac{k_{n}\beta L}{L^{2}-\beta^{2}x^{2}}=\frac{k_{n}L}{\beta\lambda_{C}^{2}z^{2}}\,.\label{derivsx}
\end{align}
Next, thanks to (\ref{recu}), (\ref{psinewdef}) and (\ref{derivsx}),
(\ref{kgcurr}) becomes 
\begin{align}
-C_{-,n}^{2}\frac{\pi}{2\beta c\hbar}\frac{1}{\cosh^{2}(\frac{\pi k_{n}}{2})}\times\nonumber \\
c^{2}\hbar\sin^{2}(\phi_{n})\frac{\partial z}{\partial x}\left[2\mathfrak{Im}(J_{-ik_{n}}(z)\frac{dJ_{ik_{n}}(z)}{dz})\right]\,.
\end{align}
In the non-relativistic limit, the last expression becomes
\begin{equation}
j_{KG}(t,x)\approx2C_{-,n}^{2}\frac{x\beta c}{L^{2}}\sin^{2}(\frac{n\pi x}{L}) \label{jnrl},
\end{equation}
 which is the expression found in \cite{mmw18}, provided $C_{-,n}^{2}\rightarrow1$. 

\section{Non-locality in the KG expanding cavity}

\subsection{Current change}

We examine here the effect of the expanding wall on the current density
at the opposite side of the cavity. In order to do so, we assume the cavity length can either
be fixed (the length is denoted by $L_0$), or in expansion, with the right wall moving at uniform
velocity as in the previous Section.  In both cases, the KG particle is prepared in the state
$\Psi_{-,1}(t=0,x)$, and the cavity length is tuned  such that\footnote{This tuning
is chosen because we require the initial
state to be the same in both cases. Note that
the preparation procedure does not have to be identical for both configurations, the requirement is that the initial state is the same.} $L(t=0)=L_0$. In the case
the cavity is expanding, we know the later current to be given by Eq. (\ref{kgcurr}). We then compare this current ($j_{e}$)
to the one ($j_{s}$) corresponding to the situation in which the
well is static. $j_{s}$ is obtained by expanding $\Psi_{-,1}(0,x)$
on the energy eigenstates of the fixed-walls cavity
\begin{equation}
\Psi_{-,1}(t=0,x)=\sum_{n}c_{n}\Phi_{-,n}(t=0,x)\,\label{cn-exp}
\end{equation}
where $\Phi_{-,n}$ is given by Eq. (\ref{kgfw}) and
\begin{align}
c_{n}=&(\Phi_{-,n},\Psi_{-,1})_{KG}&\nonumber\\
=&\int dx(\Phi_{-,n}^{*}i\hbar\partial_{t}\Psi_{-,1}+E_{n}\Phi_{-,n}^{*}\Psi_{-,1}).&\label{cn}
\end{align}
At later times, the wavefunction is given by $\Psi_{-,1}(t,x)$ when the wall is expanding and by $\sum_{n}c_{n}\Phi_{-,n}(t,x)$ 
for a stationary cavity. The resulting density currents $j_e$ and $j_s$ will therefore be different. 
Note that since we are in a non-relativistic regime, the anti-particles contribution are
not expected to be significant, as we verify below.
The next aspect to discuss is the choice of parameters and some related numerical aspects.

\subsection{Numerical aspects}

To avoid any problem of interpretation of the Klein-Gordon equation,
we work in the non-relativistic regime characterized by
\begin{equation}
\beta\ll1\quad\textrm{and}\quad z\approx\frac{L_{0}}{\lambda_{C}\beta}\gg1\,.
\end{equation}
As explained in the Appendix, we cannot numerically use the approximations
for the Bessel functions but need instead to rely on the exact forms. Since their values grow like $\cosh(\frac{\pi\nu}{2})\propto e^{\frac{\pi\nu}{2}}$
with $\nu\approx n\pi/\beta$, $\beta$ cannot be too small in order
to keep the computations tractable. For this reason, we will set $\beta=0.01$$,$
which might be somewhat higher than a typical non-relativistic case
but still abides by $\beta\ll1\text{. }$ We also need $z$ to be
large, implying $L_{0}/(\beta\lambda_{C})$ must be large. 
 
\subsection{Example}

\begin{figure}
\includegraphics[width=0.49\textwidth]{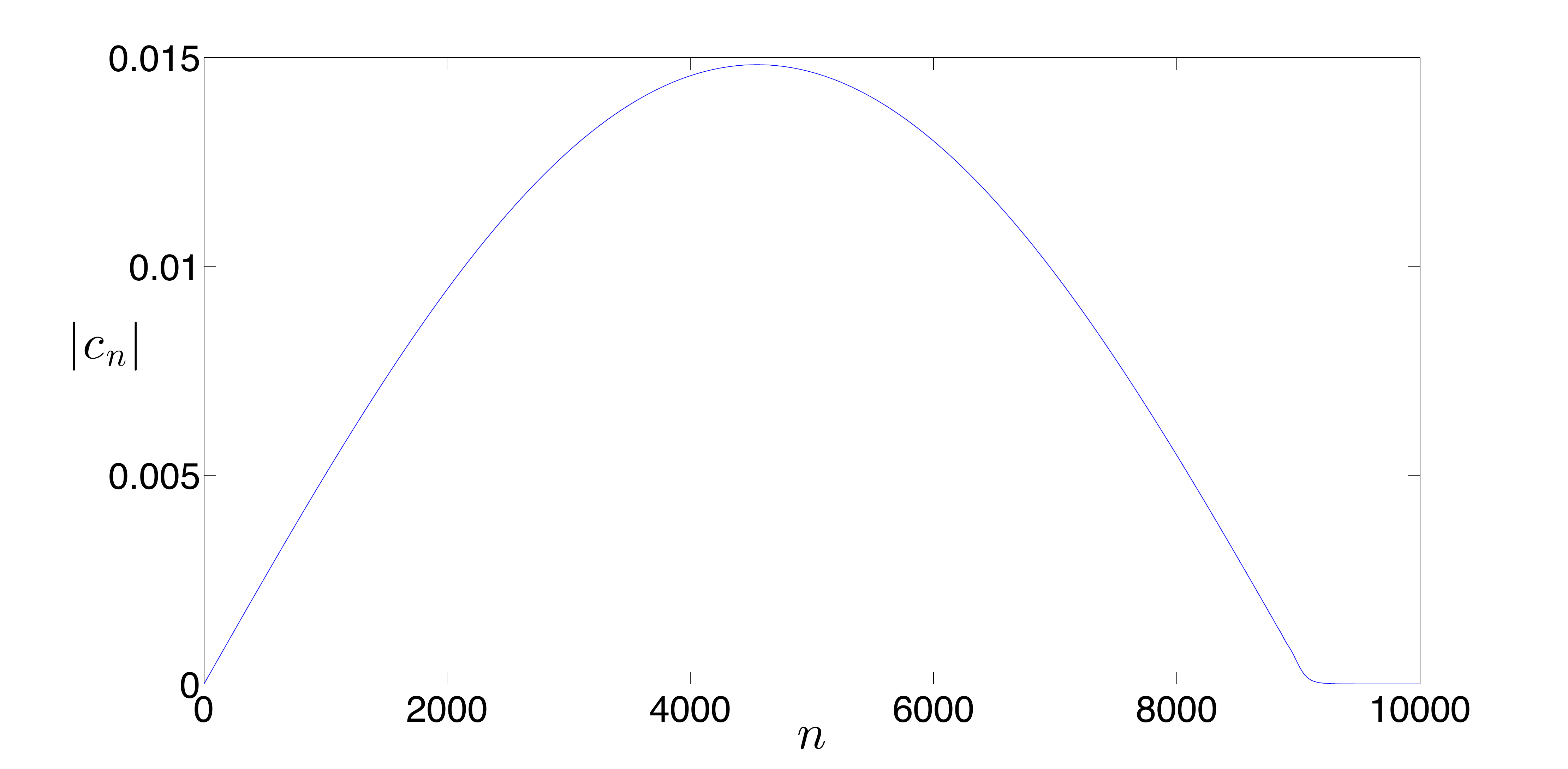} \caption{The norm of the expansion coefficients $c_{n}$ (Eq. (\ref{cn})) as a function of $n$. The maximum value is reached at $n=4551$. 
We have that $|c_{9000}/c_{4551}|=0.0312$, $|c_{10000}/c_{4551}|=2.6148\times 10^{-5}$ and $|c_{15000}/c_{4551}|=5.7616\times 10^{-7}$.}
\label{fig1} 
\end{figure}

We consider the case
\begin{equation}
m=10^{-30}\,\textrm{kg}\,\,,L_{0}=10^{-6}\,\textrm{m}\,\,\,\,\beta=0.01, \,\textrm{hence} \,\,  z\approx 10^{8}. \label{ex1}
\end{equation}
We assume that the initial state is $\Psi_{-,1}(t=0,x)$. It can be expanded over the static cavity eigenstates as per Eq. (\ref{cn}). 
We can check that, as mentioned above, the anti-particles do not contribute: the scalar products $b_n=(\Phi_{+,n},\Psi_{-,1})_{KG}$ are negligible (we have found that $|b_n|<10^{-14}$ for all $n<15000$). 
In Fig. \ref{fig1} we plot the norm of the expansion coefficients $c_n$. 

\begin{figure}
	\includegraphics[width=0.49\textwidth]{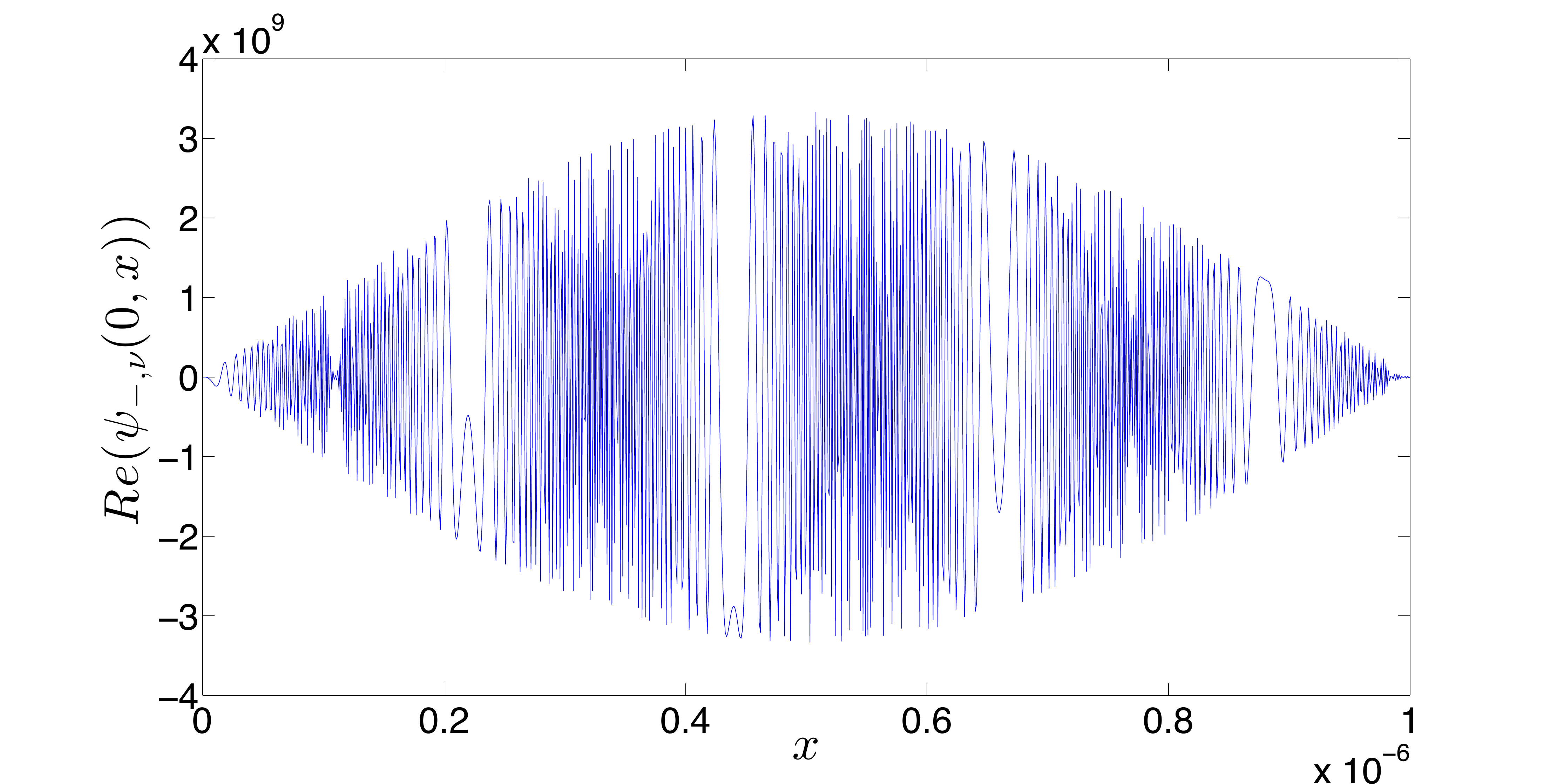} \caption{A plot of the real part of $\Psi_{-,1}(0,x)$ (Eq. (\ref{psim})),
		intended to show the large number of internal oscillations (about 500). The imaginary part behaves in a similar way.
	}
	\label{fig2}
\end{figure}

The presence probability density for $\Psi_{-,1}(0,x)$ is almost identical to the one for $\Phi_{-,1}(0,x)$ but contrary to $\Phi_{-,1}(0,x)$, $\Psi_{-,1}(0,x)$
has a large number of internal oscillations depending on the system mass and cavity
properties (see Fig \ref{fig2} for a plot of its real part). 

\begin{figure}
\includegraphics[width=0.49\textwidth]{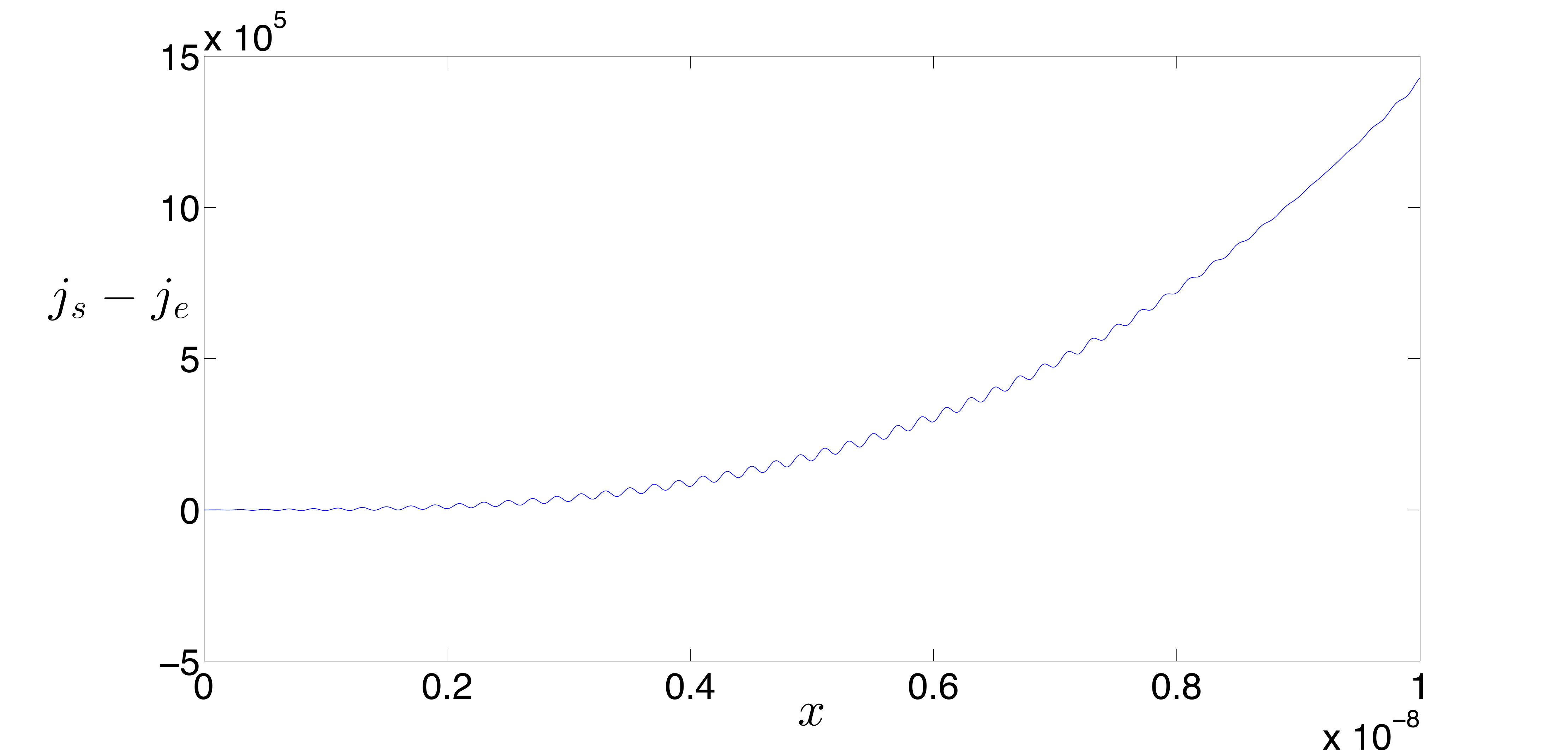} \caption{Difference between the currents $j_{s}$ and $j_{e}$ at $t=10^{-15}\,\mathrm{s}$ in a region $[0,10^{-8}\,\mathrm{m}]$ causally
disconnected from the event $(0,L_{0})$. }
\label{fig3} 
\end{figure}

Then we compare both currents in a space-time
region which is spacelike separated from the event $(t=0,x=L_{0})$. 
This illustrated in Fig. \ref{fig3} at $t=10^{-15}$ where the difference $j_s(t=10^{-15},x)-j_e(t=10^{-15},x)$ is plotted in a small region representing the leftmost  $10^{-8}/10^{-6}$ fraction of the cavity near the fixed wall.
A light signal emitted from the right wall would take at least $(10^{-6}-10^{-8})/c=3.3\times 10^{-15}$ seconds to reach this region.

For the figure, we have varied the number of steps for the numerical integration needed for the computation of the $c_n$ (the largest number of steps being $200000$) 
and $n_{max}$ (we have used $10000$ and $15000$); all runs gave almost identical curves.
The relative difference between the currents $j_e$ and $j_s$ at $x=10^{-8}\,\mathrm{m}$ is of the
order of $2\%$ while the relative difference between the box lengths at $t=0$
and $t=10^{-15}\,\mathrm{s}$ is of the order of $0.3\%$.

\section{Discussion}

The present results confirm the non-local character of the quantum
state in a relativistic setting: although the cavity is prepared in the same state at $t=0$, an observer located near the left wall ($x=0$) can discriminate as soon as $t>0$ whether the right wall is static or expanding by monitoring the local current density. This does not depend on the 
cavity size, so the right wall movement is seen to affect the current
density even if it lies in a space-like separated region (recall that the Hamiltonians
in the static and moving cases are identical except in the region beyond $x=L_0$). 

As argued in Ref. \cite{mmw18}, this effect -- if it is physical,
as we discuss below -- could in principle be used to communicate
supraluminally. Indeed, the protocol employed in \cite{mmw18}, based
on weak measurements, remains essentially the same in the present
case: a weak unitary interaction coupling the momentum of the particle
in the cavity to a pointer takes place near $x\approx0$. This unitary
interaction is immediately followed by a measurement of the position
at the same point. If the position measurement succeeds, the pointer
has shifted by a quantity proportional to the real part of the weak
value of the momentum $P^{w}$, defined by 
\begin{equation}
P^{w}=\frac{mj_{\psi}(t,x)}{\left\vert \psi(t,x)\right\vert ^{2}}-i\hbar\frac{\partial_{x}\left(\left\vert \psi(t,x)\right\vert ^{2}\right)}{2\left\vert \psi(t,x)\right\vert ^{2}}.\label{pw}
\end{equation}
This definition of the weak value is characteristic of the non-relativistic
formalism and cannot be extended straightforwardly to relativistic
wavefunctions. However we are here in the non-relativistic regime,
and we have shown above that $j_{\psi}(t,x)\approx j_{KG}(t,x)$ and
$\left\vert \psi(t,x)\right\vert ^{2}\approx\rho_{KG}(t,x)$ (see Eqs. (\ref{jnrl}) and
(\ref{psinr0}) resp.). So
from an operational point of view, the weak measurement made on a
relativistic particle in the non-relativistic regime will result in
a shift proportional to $P^{w}$ and directly depending on the current
density. 

While there is no doubt that the present relativistic model gives
rise to supraluminal communication and signaling, whether the model
is physical can be questioned. The most obvious culprit in the non-relativistic
framework was that the basis expansion akin to Eq. (\ref{cn-exp})
in principle includes states with arbitrarily high energies (leading
to arbitarily high velocities). This issue does not appear in our
relativistic framework: by definition the velocity associated with
the states in the expansion (\ref{cn-exp}) is bounded by $c$. Actually
in the numerical illustration we have given, we see from Fig. \ref{fig1}
that the expansion should include states up to $n\approx10000\text{. }$ Since
by Eq. (\ref{kgfw}) $p_{n}\approx n\pi\hbar/L_{0}=mu/\sqrt{1-\left(u/c\right)^{2}}$
where $u$ is the velocity associated to the plane wave of momentum
$p_{n}$$,$we have here $u/c\approx1.1\times10^{-2}$ for $n=10000\text{. }$

Another possible artifact could come from the breakdown of the single
particle picture that occurs when energies are sufficiently large
so that particle creation becomes possible. We do not see any obvious
reason to question the single particle picture here: first, the Klein
paradox can be avoided in our infinite well \cite{KleinP-prep};
second, as we have just seen, even the highest contributing energy
eigenstate of the fixed wall cavity ($n=10000$) yields a value for
$p/mc\approx10^{-2}$ reasonably below the particle creation threshold. A related issue
is the apparent violation of causality for a Klein-Gordon particle when the initial state is tightly localized \cite{heger}. It has been argued \cite{kimball} that this violation, that would, as in our case, give rise to signaling, is not physical, as it requires exponential localization that is not compatible with solutions of wave equations such as the Klein-Gordon equation. In the present problem the particle is not tightly localized (the cavity can be arbitrarily long), and our observation relies on solutions of the KG equation.

Since there is no obvious artifact that could account for this relativistic
non-local effect, it seems the model itself must be questioned. Indeed,
the wavefunction, be it a relativistic wavefunction, is globally defined
throughout configuration space -- here throughout the entire cavity.
If the potential changes at one end of the cavity, the wavefunction
readjusts throughout the cavity instantaneously. It is well-known \cite{transients} that a sudden change in the potential or in the boundary conditions gives rise to a transient regime. The problem here is that we are not making an approximation that could be fixed in order to account for the transient regime (other than by including an ad-hoc prescription to account for retarded effects when the wall moves). Instead, it appears that the model itself needs to be modified: for instance we might need to  model the moving wall otherwise than by a time-dependent infinite potential, for example by
a field interacting with the system through exchange particles. A
quantum field treatment of the present problem would therefore be
instructive.

\section{Conclusion}

To sum up, we have investigated a Klein-Gordon particle in an expanding
cavity. We have shown that a curious form of single particle non-locality
previously studied with the Schr\"odinger equation subsists in a relativistic
setting. Contrary to the case computed with the non-relativistic formalism,
we have not identified any obvious artifacts that could account for
the results. If we discard the possibility of this effect being physically
real (given that this form of non-locality gives rise to signaling),
our results lead to the conclusion that a relativistic model based
on a local potential affecting a wavefunction defined over an extended
region fails to capture correctly the dynamics. Investigating additional
examples as well as a quantum field based treatment would be helpful
in understanding the implications of the present results.

\medskip
\section*{Appendix}
\renewcommand{\theequation}{A-\arabic{equation}}
\setcounter{equation}{0}

\subsection*{1. KG Solutions : Asymptotic Expansion}

The functions $\tilde{J}_{\nu}(z)$ and $\tilde{Y}_{\nu}(z)$ given
by Eq. (\ref{asex}) admit the following expansion (see Eqs. (3.17)
and (3.18) of \cite{dunster}) 
\begin{eqnarray}
\tilde{J}_{\nu}(z)=\sqrt{\frac{2}{\pi z}}\bigg(\cos(z-\frac{\pi}{4})\sum_{s=0}^{s=\infty}(-1)^{s}\frac{A_{2s}(i\nu)}{z^{2s}}\nonumber \\
-\sin(z-\frac{\pi}{4})\sum_{s=0}^{s=\infty}(-1)^{s}\frac{A_{2s+1}(i\nu)}{z^{2s+1}}\bigg)\label{fullapprox1}
\end{eqnarray}
and 
\begin{eqnarray}
\tilde{Y}_{\nu}(z)=\sqrt{\frac{2}{\pi z}}\bigg(\sin(z-\frac{\pi}{4})\sum_{s=0}^{s=\infty}(-1)^{s}\frac{A_{2s}(i\nu)}{z^{2s}}\nonumber \\
+\cos(z-\frac{\pi}{4})\sum_{s=0}^{s=\infty}(-1)^{s}\frac{A_{2s+1}(i\nu)}{z^{2s+1}}\bigg)\label{fullapprox2}
\end{eqnarray}
where $A_{s}$ is defined as (see eq (4.02) of Ch. 7 in \cite{olver1974}):
\begin{equation}
A_{s}(\nu)=\frac{(4\nu^{2}-1)(4\nu^{2}-3^{2})\ldots\{4\nu^{2}-(2s-1)^{2}\}}{s!8^{s}}\,.
\end{equation}
These relations are useful as $z\rightarrow\infty$; then to lowest
order, we obtain 
\begin{align}
\tilde{J}_{\nu}(z)\approx\sqrt{\frac{2}{\pi z}}\cos(z-\frac{\pi}{4})\,\,\textrm{and}\,\,\tilde{Y}_{\nu}(z)\approx\sqrt{\frac{2}{\pi z}}\sin(z-\frac{\pi}{4})\,.\label{approx0}
\end{align}
Note that these expressions do not depend on $\nu$. Now we start
from (\ref{jsol}), we express the $J$ function in terms of $\tilde{J}$
and $\tilde{Y}$ (\ref{jrel}) and we use the approximation (\ref{approx0}).
After doing that, since $\mathrm{cosh}(\frac{\pi}{2}\nu)\approx\mathrm{sinh}(\frac{\pi}{2}\nu)$
for large $\nu$, we find that 
\begin{eqnarray}
\Psi_{J,n}\approx N_{J,n}\mathrm{cosh}(\frac{\pi}{2}k_{n})\sqrt{\frac{2\beta\lambda_{C}}{\pi\sqrt{L^{2}-\beta^{2}x^{2}}}}\times\nonumber \\
e^{+i(\frac{\sqrt{L^{2}-\beta^{2}x^{2}}}{\beta\lambda_{C}}-\frac{\pi}{4})}\mathrm{sin}(\frac{n\pi x}{L})\nonumber \\
\approx N_{J,n}\mathrm{cosh}(\frac{\pi}{2}k_{n})\sqrt{\frac{2\beta\lambda_{C}}{\pi L}}e^{i\frac{L_{0}}{\beta\lambda_{C}}}\times\nonumber \\
e^{+i\frac{mc^{2}t}{\hbar}}e^{-i\frac{\pi}{4}}\mathrm{sin}(\frac{n\pi x}{L})
\end{eqnarray}
(and similarly for $\Psi_{Y,n}$). Therefore this amounts to what
is called the ultra non-relativistic limit.
Note that it is best to use the functions $J_{i\nu}(z)\sin(\phi_{n})$
and $J_{-i\nu}(z)\sin(\phi_{n})$, also the basis used in \cite{hade},
as they form the right basis for the emergence of non-relativistic
solutions.

If we take the next terms in the series (\ref{fullapprox1}) and (\ref{fullapprox2}),
we will obtain the standard non-relativistic limit. The approximations
for (\ref{fullapprox1}) and (\ref{fullapprox2}) become 
\begin{align}
\tilde{J}_{\nu}(z)\approx & \sqrt{\frac{2}{\pi z}}\left(\cos(z-\frac{\pi}{4})+\frac{4\nu^{2}+1}{8z}\sin(z-\frac{\pi}{4})\right)\,,\label{approx1a-1}\\
\tilde{Y}_{\nu}(z)\approx & \sqrt{\frac{2}{\pi z}}\left(\sin(z-\frac{\pi}{4})-\frac{4\nu^{2}+1}{8z}\cos(z-\frac{\pi}{4})\right)\,.\label{approx1b-1}
\end{align}
Using (\ref{approx1a-1}) and (\ref{approx1b-1}) in (\ref{jrel}),
we obtain the following expression 
\begin{align}
J_{i\nu}(z) & \approx\mathrm{cosh}(\frac{\pi}{2}\nu)\sqrt{\frac{2}{\pi z}}\left(e^{i(z-\frac{\pi}{4})}[1-i\frac{4\nu^{2}+1}{8z}]\right)\nonumber \\
 & \approx\mathrm{cosh}(\frac{\pi}{2}\nu)\sqrt{\frac{2}{\pi z}}\left(e^{i(z-\frac{4\nu^{2}+1}{8z}-\frac{\pi}{4})}\right)\,, \label{approx2bis}& 
\end{align}
and the analytical solutions become Eqs. (\ref{psinr1}) and (\ref{psinr}). 

\subsection*{2. Non-relativistic limit of the KG solutions in an expanding well}

In order to show that Eq. (\ref{psinr}) reduces
to the solution of the Schr\"odinger equation (\ref{psinr0}) in the non-relativistic limit, we need to examine the imaginary exponential of Eq. (\ref{psinr}). What does the argument $z-\frac{4\nu^{2}+1}{8z}$ become in
the non-relativistic limit? The terms coming from $z$ are equal to 
\begin{equation}
\frac{L_{0}}{\lambda_{C}\beta}+\frac{mc^{2}t}{\hbar}-\frac{mvx^{2}}{2\hbar L}+\ldots\quad\,.
\end{equation}
from which we recover the term proportional to $x^{2}$ that we have
in (\ref{psinr0}). The terms coming from the other part are 
\begin{equation}
-(\frac{n^{2}\pi^{2}}{2\beta^{2}}+(\frac{1}{8}-3n^{2}\pi^{2})+\ldots)\frac{1}{z}
\end{equation}
where we have only approximated the first part, $(4\nu^{2}+1)/8\approx\frac{n^{2}\pi^{2}}{2\beta^{2}}+\frac{1}{8}-\frac{1}{3}n^{2}\pi^{2}$.
When $\beta$ is small, the dominant term is therefore $-\frac{n^{2}\pi^{2}}{2\beta^{2}z}$.
If we do a Taylor expansion of $\frac{1}{z}$ in $\beta$, we find
that $-\frac{1}{\beta^{2}z}=$ 
\begin{flalign}
&-\frac{\lambda_{C}}{L_{0}\beta} 
+\frac{c\lambda_{C}t}{L_{0}^{2}}
+\frac{\lambda_{C}}{2L_{0}^{3}}(-2c^{2}t^{2}-x^{2})\beta \nonumber \\
&+\frac{c\lambda_{C}t}{2L_{0}^{4}}(2c^{2}t^{2}+3x^{2})\beta^{2} \nonumber \\ 
&+\frac{\lambda_{C}}{8L_{0}^{5}}(-8c^{4}t^{4}-24c^{2}t^{2}x^{2}-3x^{4})\beta^{3}+(\ldots) \label{longexp}
\end{flalign}
The first term can be absorbed in the normalization factor, we factorize
$\frac{c\lambda_{C}t}{L_{0}}$ in the resulting expression ($(\ref{longexp})+\frac{\lambda_{C}}{L_{0}\beta}$)
and we keep only the terms with leading powers of $c$. Doing that
we get 
\begin{equation}
\frac{c\lambda_{C}t}{L_{0}}(\frac{1}{L_{0}}-\frac{ct}{L_{0}^{2}}\beta+\frac{c^{2}t^{2}}{L_{0}^{3}}\beta^{2}-\frac{c^{3}t^{3}}{L_{0}^{4}}\beta^{3}+(\ldots))\,,
\end{equation}
which is the Taylor expansion of 
 $ c\lambda_{C}t/(L_{0}L)= \hbar t/(m L_{0} L)$
hence the term $e^{-i\frac{\hbar n^{2}\pi^{2}t}{2mL_{0}L}}$ that
we have in (\ref{psinr0}).

\subsection*{3. Validity of the approximations and numerical simulations}
If the parameter $\frac{\nu^2}{z}$ is very small, then the approximation (\ref{approx2bis}) should be very good. 
In order to evaluate the accuracy of this approximation, we have plotted the real and imaginary parts of the quantity
\begin{equation}
\frac{J_{i\nu}(z)}{\cosh(\frac{\pi\nu}{2})\sqrt{\frac{2}{\pi z}}}-e^{i(z-\frac{4\nu^2+1}{8z}-\frac{\pi}{4})}\,,\label{ta}
\end{equation}
respectively denoted by $d_1$ and $d_2$, for a given $\nu$ ($100\pi$) and for various $z$.
Contrary to what we would expect intuitively, the approximation becomes worse for some critical value of $X=\log_{10}(z)$; at about $X=12.3$, there is a sudden increase in $d_1$ and $d_2$, from about $10^{-7}$ to $10^{-4}$. 
This is presumably due to round off errors affecting the validity of numerical routines.
Therefore, if we plan on using the approximations, we must choose our parameters (mass, initial box length and so on) in such a way as to have 
the lowest absolute error for the Bessel functions (say something of the order of $10^{-7}$).

To avoid any problem of interpretation of the Klein-Gordon equation, we work in the non-relativistic regime characterized by 
\be
\beta\ll1\quad\textrm{and}\quad z\approx\frac{L_0}{\lambda_C\beta}\gg 1\,.
\ee 
Furthermore the approximation for the Bessel function $J_{-i k_1}(z)$ (used in $\Psi_{-,1}(t,x)$) will be valid provided that 
\be
\frac{{k_1}^2}{z}\approx \frac{\pi^2}{\beta^2 z}\approx\frac{\pi^2\lambda_C}{\beta L_0}\ll 1\,.
\ee
Finally we don't want $n_{max}$ (the maximum index for the reduced basis) to be too large, otherwise the numerical integration needed to obtain the $c_n$ (see Eq. (\ref{cn})) 
would not be feasible.
To estimate $n_{max}$, we need to estimate the number of oscillations in $\Psi_{-,1}(t=0,x)$ between $x=0$ and $x=L_0$. From the expression (\ref{psinr}), we find that it is given by 
\be
\frac{z(0,0)-z(0,L_0)}{2\pi}\approx\frac{\beta L_0}{4\pi\lambda_C}.
\ee
We see that the last two conditions are antagonistic: we can't have a very good approximation of $\Psi_{-,1}(t=0,x)$ by (\ref{psinr}) unless $n_{max}$ is very large.
This explains why a numerical simulation using the approximations for the Bessel functions would be very precise only if $n_{max}$ is very large.

\end{document}